\documentclass[11pt,a4paper]{article}
\pdfoutput=1
\usepackage{jheppub}
\usepackage{amsmath}
\usepackage{amssymb}
\usepackage{amsfonts}
\usepackage{bbm}
\usepackage{graphics}
\usepackage{graphicx}
\usepackage{subfigure}
\usepackage{hhline}
\usepackage{longtable}
\usepackage{multirow}
\usepackage{dsfont}
\newcommand{\cO}{\mathcal{O}}

\newcommand{\cL}{\mathcal{L}}

\newcommand{\cH}{\mathcal{H}}

\newcommand{\cF}{\mathcal{F}}

\newcommand{\cV}{\mathcal{V}}
\newcommand{\cM}{\mathcal{M}}

\newcommand{\cZ}{\mathcal{Z}}

\def\IZ{{\mathbb Z}}
\def\IR{{\mathbb R}}

\def\IC{{\mathbb C}}
\def\IP{{\mathbb P}}

\def\IF{{\mathbb F}}
\def\IB{{\mathbb B}}

\newcommand{\hmu}{{\hat \mu}}
\newcommand{\hnu}{{\hat \nu}}
\newcommand{\hrho}{{\hat \rho}}
\newcommand{\hlambda}{{\hat \lambda}}
\newcommand{\hsigma}{{\hat \sigma}}
\newcommand{\hdelta}{{\hat \delta}}

\newcommand{\p}{\partial}

\newcommand{\be}{\begin{equation}}
\newcommand{\ee}{\end{equation}}
\newcommand{\ba}{\begin{aligned}}
\newcommand{\ea}{\end{aligned}}

\newcommand{\ben}{\begin{eqnarray}}
\newcommand{\een}{\end{eqnarray}}

\title{Five-dimensional gauge theory and compactification on a torus}
\author[a]{Babak Haghighat,}
\author[a]{Stefan Vandoren}
\affiliation[a]{Institute for Theoretical Physics and Spinoza Institute, Utrecht University, 3508 TD Utrecht, The Netherlands}
\emailAdd{B. Haghighat@uu.nl, S.J.G.Vandoren@uu.nl}

\abstract
{
We study five-dimensional minimally supersymmetric gauge theory compactified on a torus down to three dimensions, and its embedding into string/M-theory using geometric engineering. The moduli space on the Coulomb branch is hyperk\"ahler equipped with a metric with modular transformation properties. We determine the one-loop corrections to the metric and show that they can be interpreted as worldsheet and D1-brane instantons in type IIB string theory. Furthermore, we analyze instanton corrections coming from the solitonic BPS magnetic string wrapped over the torus. In particular, we show how to compute the path-integral for the zero-modes from the partition function of the M5 brane, or, using a 2d/4d correspondence, from the partition function of N=4 SYM theory on a Hirzebruch surface.
}

\begin{document}

\maketitle

%%%%%%%%%%%%%%%%%%%%%%%%%%%%%%%%%%%%%%%%%%%%%%%%%%%%%%%%%%%%%%%%%%%%%%%%%%%%%%%%%
%%%%%%%%%%%%%%%%%%%%%%%%%%%%%%%%%%%%%%%%%%%%%%%%%%%%%%%%%%%%%%%%%%%%%%%%%%%%%%%%%
\section{Introduction}

The study of the dynamics of five dimensional supersymmetric gauge theories with eight supercharges was initiated in \cite{Seiberg:1996bd}. The Coulomb branch of the theory has a rich structure and non-trivial fixed points exist at strong coupling. The BPS spectrum of the theory contains dyonic instantons \cite{Lambert:1999ua} and the solitonic magnetic string \cite{Boyarsky:2002ck}, thereby providing interesting non-perturbative phenomena.  Although non-renormalizable by power counting, these theories can be embedded into string or M-theory, either by geometric engineering \cite{Morrison:1996xf,Douglas:1996xp} or by brane constructions \cite{Aharony:1997bh}. Its maximally supersymmetric version, with sixteen supercharges, was recently studied as a candidate for the six-dimensional $(0,2)$ theory for the M5-brane \cite{Douglas:2010iu,Lambert:2010iw}.

Upon compactifying five-dimensional gauge theories on a circle of radius $R_2$, one obtains a $N=2$ $D=4$ theory. A non-perturbative solution for the low-energy effective action was proposed in \cite{Nekrasov:1996cz}, where also connections to integrable systems
were uncovered. Furthermore, at the perturbative level, resumming the one-loop contribution coming from integrating out the massive Kaluza-Klein particles, one finds an interpretation in terms of worldsheet instanton corrections in IIA
string theory \cite{Lawrence:1997jr}. All in all, there is a rich interplay between five-dimensional supersymmetric gauge theories and string theory dynamics.

In this paper, we continue the investigation of  five-dimensional gauge theories with eight supercharges,
by compactifying further down to three dimensions. We formulate the theory on $\IR^3\times T^2$, for an arbitrary torus with complex structure $\tau$ and volume ${\cal V}$. It includes the case of a rectangular torus, i.e.
the product of two circles of radii $R_1$ and $R_2$. As we will show, new phenomena appear on the Coulomb branch of the resulting three-dimensional theory. In a sense, one can view our study as a starting point to generalize the work of \cite{Gaiotto:2008cd}, where four-dimensional gauge theories were formulated on $\IR^3 \times S^1_{R_1}$. The resulting hyperk\"ahler metrics on the Coulomb branch in our case are now parametrized by the complex structure and the volume of the torus, $\tau$ and ${\cal V}$, which in the limit of $R_2\rightarrow 0$ should reduce to the metrics studied in \cite{Gaiotto:2008cd}. For generic values of $\tau$, however, one finds new hyperk\"ahler metrics which inherit modular transformation properties from the torus $T^2$, as we discuss in the main body of the paper.

To present our analysis, we choose the simplest set-up, namely we choose the gauge group to be $SU(2)$, without any additional hypermultiplets. We do not aim in this paper to find a complete non-perturbative solution of the theory, since this is beyond our abilities at present. Instead, we focus on three particular aspects of the theory, each of which is appealing in its own right:

\begin{itemize}
\item At the perturbative level, we integrate out the tower of massive Kaluza-Klein states in the one-loop approximation of the five-dimensional gauge theory on $\IR^3 \times T^2$. This produces hyperk\"ahler metrics on the Coulomb branch of the three-dimensional theory which have modular properties. Moreover, this metric is a two-parameter extension of the metric obtained in \cite{Ooguri:1996me} (see also \cite{Seiberg:1996ns,deWit:1997vg}). We verify explicity that in the limit $R_2\rightarrow 0$ we obtain back the results of \cite{Ooguri:1996me}.
In the four-dimensional limit $R_1\rightarrow \infty$, our results are again in agreement with \cite{Lawrence:1997jr}.
This analysis is done in Section 2.

\item We embed our model in string and M-theory by geometric engineering, along the lines of \cite{Morrison:1996xf,Douglas:1996xp}. The relevant set-up is M-theory compactified on $CY_3\times T^2$, which is dual to type IIB string theory. We then rederive the one-loop hyperk\"ahler metric on the Coulomb branch of the field theory from the hypermultiplet moduli space metric in type IIB \cite{RoblesLlana:2006is,Saueressig:2007dr}, in the limit where the quaternionic metric becomes hyperk\"ahler. We find that the Kaluza-Klein sum can be rewritten in terms of worldsheet and D1-brane instantons in type IIB, thereby generalizing the result of \cite{Lawrence:1997jr} to include D-brane instantons in a way that respects the modular $SL(2,\IZ)$-symmetry of the theory. The results are presented in Section 3.

\item At the non-perturbative level, we initiate the analysis of instanton corrections. We focus on the solitonic BPS magnetic string in five dimensions, wrapped over the $T^2$. This yields instanton corrections in three dimensions which correct the hyperk\"ahler metric on the Coulomb branch. From the M-theory set-up, these correspond to the euclidean M5-brane whose worldvolume is wrapped over $\IF_0 \times T^2$, where $\IF_0$ is a divisor in the Calabi-Yau. We analyze the contribution from the zero-mode sector, and argue that they can be computed from the partition function of the two-dimensional $(0,4)$ CFT obtained by wrapping the M5-brane on $\IF_0$. In type IIB theory, these instantons correspond to $D3$ branes wrapping $\IF_0$, whose dynamics is described by four-dimensional $N=4$ super Yang-Mills theory on $\IF_0$ with gauge coupling constant identified with the complex structure of the torus, $\tau$. Invoking a 2d/4d correspondence, we conjecture that the partition function of the 4d SYM theory computes the zero mode contribution of the instanton corrections as well. We demonstrate this explicitly for the case when the magnetic string has magnetic charge equal to one. The derivation can be found in Section 4.

\end{itemize}

Our investigations, though concrete, do not provide a complete description of the non-perturbative structure of the five-dimensional theory on $\IR^3\times T^2$. In the final section, we therefore outline some directions for future research.

\section{Field theory description}

In this section we first review five-dimensional $SU(2)$ gauge theories and their low-energy effective action in the broken phase of the gauge theory.
Then we pass over to compactify the theory on the two-torus yielding an $N=4$ supersymmetric theory in three dimensions whose classical bosonic action we derive.
This will define a sigma-model where the target space is endowed with a hyperk\"ahler metric. Last but not least we will study one-loop perturbative quantum corrections to this metric.

\subsection{Five-dimensional $SU(2)$ gauge theories}
\label{5Dfieldth}

In the following we consider five-dimensional $SU(2)$ gauge theories with $N=1$ supersymmetry. The dynamical fields are
those of an on-shell five-dimensional vector-multiplet given by a real scalar $\sigma$, a vector field $A_{\hmu}$ and a Dirac spinor
$\lambda$ in the adjoint of $SU(2)$. The signature of the metric is $(+,-,-,-,-)$ and the five spacetime dimensions are denoted by hatted Greek indices $\hmu = 0,1, \cdots, 4$. Denoting the bare five-dimensional coupling constant by $g_{5,0}$ the on-shell Lagrangian can be written as follows
\begin{equation} \label{nonabelian}
    \cL^{(1,4)} = \frac{1}{g_{5,0}^2}\textrm{Tr} \left\{- \frac{1}{4} F_{\hmu \hnu} F^{\hmu \hnu}
                  - \frac{1}{2} D_{\hmu}\sigma D^{\hmu}\sigma - \frac{1}{2} \bar{\lambda} \Gamma^{\hmu} D_{\hmu} \lambda
                  - \frac{1}{2} \bar{\lambda} \left[\sigma,\lambda\right] \right\}.
\end{equation}
Here the field strength is given by
\begin{equation}
    F_{\hmu \hnu} = (\p_{\hmu} A_{\hnu} - \p_{\hnu} A_{\hmu}) - \left[A^c_{\hnu}, A^d_{\hmu} \right],
\end{equation}
and the covariant derivative acts as
\begin{equation}
    D_{\hmu} X = i \p_{\hmu} X + \left[A_{\hmu}, X \right],
\end{equation}
where all quantities take value in the Lie algebra, i.e. $X = X^a \tau_a$, where the $\tau_a$, $a=1,2,3$, are the Pauli matrices.

\subsubsection*{Spontaneous breaking to $U(1)$}
\label{U1gaugeth}

The non-abelian theory described by (\ref{nonabelian}) can be spontaneously broken to $U(1)$ by giving a vev to the scalar field
\begin{equation}
    \sigma = \phi \tau_3 + \delta \sigma,
\end{equation}
with $\phi>0$\footnote{Here we have taken into account the action of the Weyl group $\IZ_2$.}. Thus, the Coulomb branch is parameterized by $\phi \in \IR^+$. The massive particles will now consist of $W$-bosons
\begin{equation}
    A^+_{\hmu} = A^1_{\hmu} - i A^2_{\hmu}~, \quad A^-_{\hmu} = A^1_{\hmu}+i A^2_{\hmu}~,
\end{equation}
and their fermionic superpartners
\begin{equation}
    \lambda^+ = \lambda^1 - i \lambda^2, \quad \lambda^- = \lambda^1 + i \lambda^2.
\end{equation}
Together they form an on-shell massive five-dimensional vector multiplet of mass $2\phi$. The fermionic and bosonic degrees of freedom
accordingly add up to $4 + 4 = 8$. Note that the $\delta \sigma^{\pm} = \delta \sigma^1 \mp i \delta \sigma^2$ remain massless
as they are the Goldstone-modes of the symmetry breaking mechanism.

Going to the low energy effective field theory by integrating out massive multiplets one arrives at a supersymmetric $U(1)$ gauge theory.
Such theories are controlled by a prepotential \cite{Gunaydin:1983bi,Cortes:2003zd}
\begin{equation} \label{prepotential}
    \cF = \frac{a_0}{2} \phi^2 + \frac{\kappa}{6} \phi^3,
\end{equation}
for some real constants $a_0$, $\kappa$ $\in \IR$. The effective gauge coupling is given by the second derivative of $\cF$:
\begin{equation}
    a(\phi) = \p_{\phi}^2 \cF = a_0 + \kappa \phi.
\end{equation}
Even if the constant $\kappa$ in (\ref{prepotential}) is zero classically it can be created at one-loop in the quantum theory \cite{Witten:1996qb}.
For a nonzero $\kappa$ supersymmetry \cite{Gunaydin:1983bi,Cortes:2003zd} requires the presence of a Chern-Simons term
\begin{equation} \label{chernsimons}
    \frac{\kappa}{24 \pi^2} A \wedge F \wedge F.
\end{equation}
For $SU(2)$ gauge theories with $N_f$ "quarks" which are hypermultiplets in the two-dimensional representation of the gauge group
a one-loop computation yields the following gauge coupling \cite{Seiberg:1996bd}
\begin{equation} \label{gaugecoupling}
    a(\phi) = \frac{1}{g(\phi)^2} = \frac{1}{g_{5,0}^2} + 16 \phi - \sum_{i=1}^{N_f} |\phi - m_i| - \sum_{i=1}^{N_f} |\phi + m_i|~.
\end{equation}
Here, $\frac{1}{g_{5,0}^2}$ is the bare coupling and the term $16 \phi$ comes from integrating out the $W$-bosons each giving a contribution which is the cube of the charge and thus equals $8$.  The $m_i$ correspond to the masses of the hypermultiplets. As we have no hypermultiplets in our setup
the two last terms in (\ref{gaugecoupling}) fall away, hence we have
\begin{equation}
    a_0 = \frac{1}{g_{5,0}^2}~, \quad \kappa = 16.
\end{equation}
Thus, the bosonic part of the Lagrangian of the effective $5$-dimensional $U(1)$ theory  is
\begin{equation} \label{5daction}
    \cL_{bos}^{(1,4)} = \left(-\frac{1}{4} F_{\hmu \hnu} F^{\hmu \hnu} - \frac{1}{2} \p_{\hmu} \sigma \p^{\hmu} \sigma\right)a(\sigma)
                        - \frac{\kappa}{24\pi^2} \epsilon^{\hmu \hnu \hlambda \hrho \hsigma} A_{\hmu} F_{\hnu \hlambda} F_{\hrho \hsigma}~,
\end{equation}
where, as usual, the vev $\phi$ is promoted to a dynamical field, denoted by $\frac{1}{2}\sigma$.

\subsubsection*{BPS states}
\label{5Dbps}

The massive spectrum of five-dimensional supersymmetric $SU(2)$ gauge theories in the abelian phase includes BPS saturated states \cite{Seiberg:1996bd}. First of all such states include particles which are electrically charged and their masses and central charges are given by
\begin{equation}
    \frac{m_e}{\sqrt{2}} = Z_e = n_e \phi, \quad n_e \equiv \int_{S^3_{\infty}} \ast F.
\end{equation}
Furthermore, the vector multiplet can be dualized to become a tensor multiplet including a two-form
\footnote{Strictly speaking the dualization cannot be carried out at the level of the Lagrangian once the Chern-Simons term
is included. However, the solitonic string-solution will still be present which justifies the inclusion of the two-form.} gauge field $B_{\mu \nu}$ and a dual scalar
\begin{equation} \label{stringtension}
    \phi_D = \frac{\p \cF}{\p \phi}(\phi) = \frac{1}{g_{5,0}^2}\phi + \frac{\kappa}{2} \phi^2.
\end{equation}
The theory then admits solitonic string-like objects which are magnetically charged under $B_{\mu \nu}$. In the BPS saturated case the tension and central charge of these strings is given by
\begin{equation}\label{tension-string}
    \frac{T}{\sqrt{2}} = Z_m = n_m \phi_D, \quad n_m \equiv \frac{1}{4\pi} \int_{S^2_{\infty}} \epsilon_{abc} \sigma^a d \sigma^b \wedge d \sigma^c,
\end{equation}
where $\sigma$ is a monopole configuration with $\sum_a \left(\sigma^a\right)^2 = \phi^2$ at $S^2_{\infty}$. For more details on the field configuration of this string
monopole we refer to \cite{Boyarsky:2002ck}.
A third class of BPS states consists of four-dimensional instantons lifted to solitons in $4+1$ dimensions. They were called dyonic instantons in \cite{Lambert:1999ua}. Their mass is given by $m_I = \frac{|n_I|}{g_{5,0}^2}$
where $n_I$ is the four-dimensional instanton number. However, due to the one-loop correction (\ref{chernsimons}) these instantons become electrically charged under the
$U(1)$ gauge field and their total contribution to the central charge can be written as
\begin{equation}\label{central-charge-DI}
    Z_I = \kappa \phi |n_I| + \frac{|n_I|}{g_{5,0}^2}, \quad n_I = \frac{1}{8 \pi^2} \int_{\IR^4} \textrm{Tr}~F \wedge F.
\end{equation}
The above states are important to determine the 3D effective action, as they have to be integrated out when compactifying on $T^2$.

\subsection{Dimensional reduction on $T^2$}
\label{torusreduction}

In this section we compactify the low-energy effective Lagrangian on $T^2$ along the directions $\hmu=3,4$. We will henceforth
use Greek indices for the resulting three-dimensional Minkowski space-time and Latin indices $i=1,2$ for the directions along the $T^2$.
In order to compactify the theory on $T^2$ we need to define the normalization of the gauge fields along the compact directions. This is done
by demanding invariance under large gauge transformations
\begin{equation}
    \oint_{S^1_i} 2A_i \mapsto \oint_{S^1_i} 2A_i + 2\pi.
\end{equation}
The Wilson line variables
\begin{equation}
    \varphi_1 \equiv 2 \oint_{S^1_3} A_3, \quad \varphi_2 \equiv 2 \oint_{S^1_4} A_4, \label{phidef}
\end{equation}
are therefore periodic variables and parameterize a torus
\begin{equation}
    (\varphi_1,\varphi_2) \in \Gamma \otimes_{\IZ}(\IR/2\pi \IZ),
\end{equation}
with $\Gamma \simeq \IZ^{2}$. Under the $SL(2,\IZ)$-symmetry of $T^2$ the $\varphi_i$ transform as follows
\begin{eqnarray}
    \left(\begin{array}{c} \varphi_2 \\ \varphi_1 \end{array} \right)
        & \mapsto & \left(\begin{array}{cc} a & b \\ c & d \end{array}\right)\left(\begin{array}{c} \varphi_2 \\ \varphi_1 \end{array} \right),
\end{eqnarray}
where $\left(\begin{array}{cc}a & b\\ c & d\end{array}\right) \in SL(2,\IZ)$.

Now let's look at the reduction of the first term in (\ref{5daction}). As a first step introduce the metric on $\IR^3\times T^2$
\begin{equation} \label{5dmetric}
    g_{\hmu \hnu} = \left(
        \begin{array}{cc}
            g_{ij} & 0      \\
            0      & \eta_{\mu \nu}
        \end{array}
    \right),
\end{equation}
where
\begin{equation} \label{Tmetric}
    g_{ij} = \frac{\cV}{\tau_2}\left(\begin{array}{cc}1 & \tau_1 \\ \tau_1 &  |\tau|^2 \end{array} \right),
\end{equation}
with $\cV$ being the volume of the $T^2$ and $\tau = \tau_1 + i \tau_2$ the complex structure. Using that $\partial_{i} A_{\hmu} = 0$
we obtain
\begin{equation}
    \int_{T^2} \sqrt{g} F_{\hmu \hnu} F_{\hdelta \hlambda} g^{\hdelta \hmu} g^{\hlambda \hnu} a(\sigma)
    = \cV F_{\mu\nu} F^{\mu\nu} a(\sigma) + \frac{2}{\tau_2} \p_{\mu} z \p^{\mu} \bar{z}~a(\sigma),
\end{equation}
where
\begin{equation} \label{zdef}
    g = \textrm{det}g_{ij}, \quad z = \varphi_2 - \tau \varphi_1, \quad \bar{z} = \varphi_2 - \bar{\tau} \varphi_1.
\end{equation}
Under the $SL(2,\IZ)$-symmetries of the $T^2$, $\tau$ does not stay invariant and the orbits generated by its transformations define
distinct equivalence classes. Therefore, $\tau$ is not valued in the upper half plane $\cH = \{x \in \IC | \textrm{Im}(x) > 0\}$ but rather takes values in the fundamental domain $\cH/PSL(2,\IZ)$. More precisely, we have the following transformation rules
\begin{equation} \label{sl2z}
    z \mapsto \frac{z}{c\tau + d}, \quad \bar{z} \mapsto \frac{\bar{z}}{c \bar{\tau} + d},
    \quad \tau \mapsto \frac{a\tau + b}{c\tau + d},\quad \left(\begin{array}{cc}a & b \\c & d\end{array}\right) \in SL(2,\IZ).
\end{equation}
The kinetic terms of the Lagrangian only stay invariant under the above transformations if $\tau$ is understood as an element of the
fundamental domain, otherwise not. Let us elaborate on the remaining terms in the Lagrangian (\ref{5daction}) to extract the full symmetry group.
Under compactification on the torus they reduce as follows:
\begin{eqnarray}
    -\frac{1}{2} \p_{\hmu}\sigma \p^{\hmu} \sigma
        & \rightarrow & -\frac{1}{2} \p_{\mu} \sigma \p^{\mu} \sigma~, \\
    -\frac{1}{24}\epsilon^{\hmu \hnu \hlambda \hrho \hsigma} A_{\hmu} F_{\hnu \hlambda} F_{\hrho \hsigma}
        & \rightarrow & \frac{1}{2} \epsilon^{\mu\nu\lambda} \varphi_1 \p_{\mu} \varphi_2 F_{\nu\lambda}.
\end{eqnarray}
Thus altogether we arrive at the bosonic part of the Lagrangian
\begin{eqnarray} \label{3dlag}
    \cL & = & \int_{T^2} \sqrt{g} \cL^{(1,4)}_{bos} \nonumber \\
    ~                 & = & \left(-\frac{\cV}{4}F_{\mu\nu} F^{\mu\nu}  - \frac{1}{2\tau_2} \p_{\mu} z \p^{\mu} \bar{z}
                            -\frac{\cV}{2} \p_{\mu} \sigma \p^{\mu} \sigma\right) a(\sigma) \nonumber \\
    ~                 & ~ & + \frac{\cV}{2\pi^2} \kappa ~\epsilon^{\mu\nu\lambda}\varphi_1 \p_{\mu} \varphi_2 F_{\nu\lambda}.
\end{eqnarray}
In the above expression $\cV a(\sigma)$ starts with $\frac{\cV}{g_{5,0}^2} + \cdots$, so we can read off the bare coupling constant of the three-dimensional
field theory,
\begin{equation}
    \frac{1}{g_{3,0}^2} = \frac{\cV}{g_{5,0}^2}.
\end{equation}
At first glance this Lagrangian is not invariant under the interchange of $\varphi_1$ and $\varphi_2$ as it should be. However,
this interchange is an element of $SL(2,\IZ)$ and therefore showing invariance under $SL(2,\IZ)$ will solve the issue. Indeed
one can easily show that the above Lagrangian is invariant under the transformations (\ref{sl2z}) up to a surface-term proportional to
\begin{equation}
    \p_{\mu} \epsilon^{\mu \nu \lambda}\left[\frac{ac}{2} ~\varphi_1 \varphi_1 F_{\nu\lambda} + \frac{bd}{2} ~\varphi_2 \varphi_2 F_{\nu\lambda}
    + bc~\varphi_2 \varphi_1 F_{\nu\lambda} \right].
\end{equation}
This is not yet the full symmetry group. There are additional continuous isometries of the torus which after compactification
to three dimensions should manifest themselves as translations of the fields $\varphi_i$. Under such translations we pick up the following
surface terms
\begin{eqnarray}
    \varphi_1 \mapsto \varphi_1 + \alpha & : & \quad \cL \mapsto \cL + \frac{\cV}{2\pi^2}\kappa \alpha \epsilon^{\mu \nu \lambda} \p_{\mu} \left(\varphi_2 F_{\nu \lambda}\right)~, \label{tr1}\\
    \varphi_2 \mapsto \varphi_2 + \beta  & : & \quad \cL \mapsto \cL - \frac{\cV}{2\pi^2}\kappa \beta \epsilon^{\mu \nu \lambda} \p_{\mu} \left(\varphi_1 F_{\nu \lambda}\right) \label{tr2}.
\end{eqnarray}
The Lagrangian (\ref{3dlag}) describes the dynamics of a three-dimensional tensor-multiplet as it contains the gauge field $A_{\mu}$. In order to
switch to the hypermultiplet picture we have to dualize its action. For this task we introduce a Lagrange-multiplier $\lambda$ which modifies the action to the form
\begin{eqnarray} \label{laction}
    \cL_{\lambda,F} & = & \left(-\frac{\cV}{4}F_{\mu\nu} F^{\mu\nu} - \frac{1}{2\tau_2} \p_{\mu} z \p^{\mu} \bar{z}
                            -\frac{\cV}{2} \p_{\mu} \sigma \p^{\mu} \sigma\right) a(\sigma) \nonumber \\
    ~                 & ~ & + \frac{\cV}{2\pi^2} \kappa ~\epsilon^{\mu\nu\delta}\varphi_1 \p_{\mu} \varphi_2 F_{\nu\delta}  + \frac{\lambda}{8\pi} \p_{\mu} F_{\nu\delta} \epsilon^{\mu\nu\delta}.
\end{eqnarray}
Here, it is important to note, that in order for the action to be invariant under the translations (\ref{tr1}), (\ref{tr2}) $\lambda$ has to transform as follows
\begin{equation} \label{lambdatrafo}
    \lambda \mapsto \lambda
    + \frac{4\cV}{\pi} \kappa ~(\alpha \varphi_2 - \beta \varphi_1).
\end{equation}\
Hence the torus action acts non-trivially on the Lagrange-multiplier.
Now we integrate out the gauge field by using its equation of motion. Varying with respect to $F_{\mu\nu}$ we get
\begin{equation}
    \frac{\delta \cL_{\lambda,F}}{\delta F_{\mu\nu}}
    = -\frac{\cV}{2} F^{\mu\nu} a(\sigma) + \frac{\cV}{2\pi^2} \kappa ~\epsilon^{\delta\mu\nu} \varphi_1 \p_{\delta} \varphi_2
      -  \frac{\p_{\delta} \lambda}{8\pi} \epsilon^{\delta\mu\nu},
\end{equation}
and therefore
\begin{equation}
    F^{\mu\nu} = \frac{2}{\cV} a(\sigma)^{-1}\left[\frac{\cV}{2\pi^2}\kappa~\epsilon^{\delta\mu\nu}\varphi_1\p_{\delta} \varphi_2  -  \frac{\p_{\delta} \lambda}{8\pi} \epsilon^{\delta\mu\nu}\right].
\end{equation}
Inserting this back into (\ref{laction}) we arrive at
\begin{eqnarray}
    \cL_{\lambda}
        & = & \left(- \frac{1}{\tau_2 \cV} \p_{\mu} z \p^{\mu} \bar{z}
              - \p_{\mu}\sigma \p^{\mu} \sigma \right) \frac{\cV}{2} a(\sigma) \nonumber \\
    ~   & ~ & + \frac{2}{a(\sigma) \cV}\left[\frac{\cV}{2\pi^2}\kappa~ \varphi_1 \p_{\delta} \varphi_2  -   \frac{\p_{\delta}\lambda}{8\pi}\right]
                                \left[\frac{\cV}{2\pi^2}\kappa~ \varphi_1 \p^{\delta} \varphi_2  -   \frac{\p^{\delta}\lambda}{8\pi}\right]~, \label{Lclass}
\end{eqnarray}
where we remark that $z$ is a function of the $\varphi_i$ as stated in (\ref{zdef}).

\subsubsection*{The structure of the hyperk\"ahler metric}

The Lagrangian (\ref{Lclass}) takes the form of a non-linear sigma model with $N=4$ supersymmetry in three dimensions. The resulting metric must therefore
be hyperk\"ahler, and for the case of (\ref{Lclass}), it fits into the class of metrics described in \cite{Pederson}, see also \cite{Hitchin:1986ea,DeJaegher:1997ka}. In general, $4n$ dimensional hyperk\"ahler metrics with $n$ commuting isometries can be written as \cite{Pederson}
\begin{equation} \label{hkmetric}
    ds^2 = U_{IJ}(x) d\vec{x}^I \cdot d\vec{x}^J + (U^{-1}(x))^{IJ}(d\varrho_I + \vec{W}_{IK}(x)\cdot d\vec{x}^K)
       (d\varrho_J + \vec{W}_{JL}(x)\cdot d\vec{x}^L).
\end{equation}
Compared to our case the isometry is given by a shift in $\lambda$, which is preserved in perturbation theory. We find that $I$ and $J$ can be set to $1$ or omitted as there is only one hypermultiplet. Furthermore, we have
\begin{eqnarray} \label{su2hkmetric}
    x_1 & = & \varphi_1, \quad x_2 = \varphi_2, \quad x_3 = \sigma, \quad \varrho = - \frac{\lambda}{8\pi}, \\
    U_{class}   & = & \frac{\cV}{2} a(\sigma), \quad \vec{W}_{class} = (v^j g_{ji}, 0), \quad \vec{v} = \left(0, -\frac{\cV}{2\pi^2}\kappa~\varphi_1\right). \label{UWclass}
\end{eqnarray}
$U$ and $W$ are related through the following equation
\begin{equation}
    \vec{\nabla} U = \vec{\nabla} \times \vec{W},
\end{equation}
which is automatically satisfied for the classical functions (\ref{UWclass}) where we have used that $a(\sigma) = a_0 + \frac{\kappa}{2} \sigma$.
Note that the dot-product is not the Euclidean one but rather given by
\begin{equation} \label{dotmetric}
    \vec{a} \cdot \vec{b} = \left(\vec{a}\right)^T \left(\begin{array}{cc}g^{-1} & 0 \\ 0 & 1\end{array}\right) \vec{b}~,
\end{equation}
where $g^{-1}$ is the metric $g^{ij}$ defined in (\ref{Tmetric})\footnote{The metric (\ref{dotmetric}) can be diagonalized to obtain the Euclidean metric used in reference \cite{DeJaegher:1997ka}.}.
This is due to the fact that there exists an $SL(2,\IZ)$-action which leaves the Lagrangian (\ref{Lclass}) invariant. The isometry
group of the metric is given by constant shifts in the fields $(x_1,x_2,\varrho)$. In the quantum theory (see section \ref{qc}) the continuous shifts
in $x_1, x_2$ become discrete. Furthermore, $\varrho$ becomes periodic due to three-dimensional instantons \cite{Polyakov}. As noted in \cite{Dorey:1997ij} this works as follows. The euclidean Lagrangian (\ref{laction}) induces the term
\begin{equation}
    S_{top} = i\frac{\langle \lambda\rangle}{8\pi} \int d^3 x \epsilon^{\mu \nu \rho} \p_{\mu} F_{\nu \rho}
\end{equation}
into the three-dimensional action. As the 3D instanton topological charge is given by
\begin{equation}
    k = \frac{1}{8\pi} \int d^3 x \epsilon^{\mu \nu \rho} \p_{\mu} F_{\nu \rho} \in \IZ~,
\end{equation}
we see that $\lambda$ is periodic with period $2\pi$. As explained in \cite{Polyakov} instantons in three dimensions are magnetic monopoles in four
dimensions. Uplifting this to five dimensions yields the magnetic string. Indeed, the dual scalar $\lambda$ can be directly obtained from five
dimensions via
\begin{equation}
    \lambda = \int_{T^2} B_{ij} dx^i \wedge dx^j.
\end{equation}
We know that in five dimensions the objects which are charged under $B_{\mu \nu}$ are the solitonic strings. Thus we conclude that the
instantons in three dimensions to which $\lambda$ couples are obtained by wrapping the world-volume of the five-dimensional string on the $T^2$.

Altogether we see that the resulting hyperk\"ahler metric is a fibration $\Sigma$ over $\IR$ where locally $\Sigma \simeq T^2_{\varphi} \times S^1_{\lambda}$ and $\IR$ is parameterized by the field $\sigma$. We can exhibit even more structure. Note from (\ref{lambdatrafo}) that the shift symmetry $\varphi_1 \mapsto \varphi_1 + \alpha$ with $\alpha$
constant is only a symmetry if $\lambda$ simultaneously transforms as $\lambda \mapsto \lambda + \frac{4\cV}{\pi}\kappa \alpha \varphi_2$. Thus we have identified
$\frac{4\cV}{\pi}\kappa \varphi_1 d \varphi_2$ as the connection for the circle bundle $S^1_{\lambda}$ over $T^2_{\varphi}$.
Let us furthermore define
\begin{eqnarray}
    T_1 & : & \varphi_1 \mapsto \varphi_1 + \alpha, \nonumber \\
    T_2 & : & \varphi_2 \mapsto \varphi_2 + \beta, \\
    T_3 & : & \lambda~ \mapsto \lambda + \gamma. \nonumber
\end{eqnarray}
Then the $T_i$ form a Heisenberg-algebra as can be seen by computing their commutators:
\begin{equation}
    \left[T_2,T_3\right] = 0, \quad \left[T_1,T_2\right] = T_3~, \quad
    \left[T_1,T_3\right] = 0.
\end{equation}
The resulting structure is similar to the quaternionic hypermultiplet moduli space of type IIA string compactifications on $CY_3$-manifolds
\cite{Alexandrov:2010np,Alexandrov:2010ca,Alexandrov:2008gh,Alexandrov:2006hx,Davidse:2004gg}, where NS5-brane instantons play the role of our magnetic string instantons, and the RR-scalars parametrize the torus of the intermediate Jacobian of the $CY_3$. Topologically, the hyperk\"ahler space can therefore be depicted as in figure 1.

\begin{figure}[h]
\label{figure:torusfib}
\center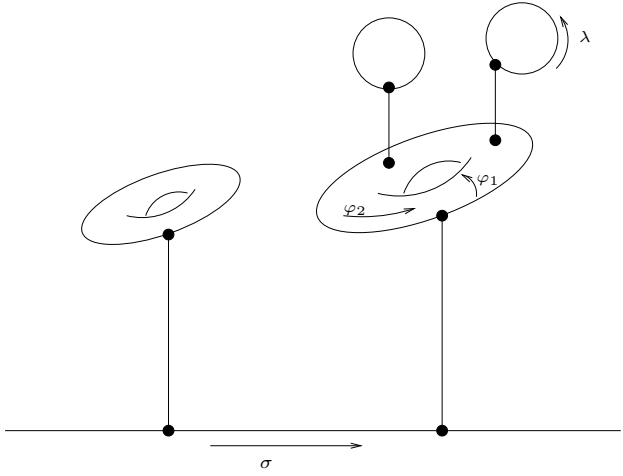
\caption{Toric Hyperk\"ahler space}
\end{figure}

\subsection{Integrating out Kaluza-Klein states}
\label{qc}

The metric derived in section \ref{torusreduction} is only valid in the limit where the volume of the torus is large.
Here we will derive quantum corrections to the gauge kinetic function in three dimensions corresponding to the function $U$ in (\ref{hkmetric})
which become more important when the volume of the $T^2$ is decreased.
Before compactifying on a torus however, we will first review the case of compactification on a circle which was already studied in
\cite{Nekrasov:1996cz, Lawrence:1997jr}.

In order to compute the effective action at one-loop let us have a look at terms quadratic in $\lambda^{\pm}$ in
the Lagrangian (\ref{nonabelian}):
\begin{equation} \label{fermprop}
    \bar{\lambda}^{-} \left[\Gamma^{\hmu} (i \p_{\hmu} + 2 A^3_{\hmu}) + 2\phi + 2 \delta \sigma^3\right]\lambda^+ +
    \bar{\lambda}^{+} \left[\Gamma^{\hmu} (i \p_{\hmu} - 2 A^3_{\hmu}) - 2\phi - 2 \delta \sigma^3 \right]\lambda^-.
\end{equation}
In the following we will use the above expression to extract propagators for the gauge theory on the geometries $\IR^4 \times S^1$ and $\IR^3 \times T^2$.
The extraction of the propagators for the massive gauge bosons requires a gauge fixing procedure which we omit here for clarity of presentation. Due to supersymmetry, they yield the same final expression for the hyperk\"ahler function $U$ with a different numerical constant which we fix after equation \eqref{TN}.

\subsubsection*{From 5D to 4D}
\label{5d4d}

The $5D$ field theory is in the Coulomb branch of $SU(2)$ parametrized by the vev of the scalar $\phi \geq 0$.
By compactifying the $5D$ theory on $S^1$ with radius $R$ the five-dimensional vector-field will split according to
\begin{equation}
    A_{\hmu} \rightarrow \left\{A_{\mu},~ A_{4}\right\},
\end{equation}
and the component $A_4$ will combine with the five-dimensional scalar $\sigma$ to form a complex four-dimensional scalar.
To obtain the one-loop correction to the gauge coupling by taking into account all Kaluza-Klein modes of the massive vector-multiplet
along the $S^1$, we have to extract the propagators of the massive fields. It suffices to do this for the fermions.
The massive gauge bosons sit in the same multiplet and yield similar expressions.
For that we rewrite (\ref{fermprop}) for the geometry $\IR^{1,3} \times S^1$:
\begin{eqnarray}
    ~ & ~ & \sum_{m,n} \bar{\lambda}_m e^{i \frac{m}{R} x_4} \left[\Gamma^{\hmu} (i \p_{\hmu} + 2 A_{\hmu}) + \sigma \right]\lambda_n e^{-i \frac{n}{R} x_4} \nonumber\\
    ~ & = & \sum_{m,n} \bar{\lambda}_m e^{i \frac{m}{R} x_4} \left[\Gamma^{\mu} (i \p_{\mu} + 2 A_{\mu}) + \Gamma^{4} (\frac{n}{R} + 2 A_4) + \sigma \right]
            \lambda_n e^{-i \frac{n}{R} x_4},
\end{eqnarray}
where $\mu=0,1,\cdots,3$. After integration with respect to $x_4$ we obtain the propagators in momentum space
\begin{equation} \label{4dprop}
    \langle \lambda_n \bar{\lambda}_n \rangle
    = \frac{1}{\Gamma^{\mu} k_{\mu} + \Gamma^4 (\frac{n}{R} + 2 A_4) + \sigma}
    =   \frac{\Gamma^{\mu} k_{\mu} + \Gamma^{4} (\frac{n}{R} + 2 A_4) - \sigma}{k^2 + \frac{1}{R^2} (n + 2 R A_4)^2 - \sigma^2}.
\end{equation}
Now, standard quantum field theory techniques imply that the one-loop amplitude with two external gauge bosons at polarizations $\alpha$ and $\beta$
obtained by using the propagator (\ref{4dprop}) is, at second order in the external momentum $p$, proportional to
\begin{equation}
    (p_{\alpha} p_{\beta} - \eta_{\alpha \beta} p^2)
    \sum_{n} \int \frac{d^4 k}{(2\pi)^4} \frac{1}{\left[k^2 + \frac{1}{R^2} (n + 2 R A_4)^2 + \sigma^2\right]^2}~,
\end{equation}
where the momentum integral is now over $\IR^4$. The integrand can be rewritten using the formula
\begin{equation}
    \frac{1}{X^2} = \int_{0}^{\infty} ds s e^{-s X},
\end{equation}
and the whole integral becomes
\begin{equation}
    \int \frac{d^4 k}{(2\pi)^4} \int_{0}^{\infty} ds s e^{-s \left[k^2 + \frac{1}{R^2} (n + 2 R A_4)^2 + \sigma^2\right]}
    = \frac{1}{16 \pi^2} \int_{0}^{\infty} \frac{ds}{s} e^{-s \left[\frac{1}{R^2} (n + 2 R A_4)^2 + \sigma^2\right]}~.
\end{equation}
Poisson resummation with respect to $n$ further gives\footnote{We have left out the mode $n=0$. It yields a diverging contribution and needs to be
substracted in a proper renormalization scheme. In fact, this divergence already occurs in the decompactification limit $R\rightarrow \infty$.}
\begin{eqnarray}
    ~ & ~ & \frac{1}{16 \pi^2} \sum_{n\neq 0} e^{-2\pi i n R 2 A_4} \int_{0}^{\infty} \frac{ds}{s^{3/2}} e^{-s \sigma} \sqrt{\pi} R e^{-\pi^2 n^2 \frac{R^2}{s^2}}\
            \nonumber \\
    ~ & = & \frac{1}{16 \pi^2} \sum_{n\neq 0} \frac{1}{|n|} e^{-2 \pi i R n 2 A_4 - 2 \pi |n| R \sigma},
\end{eqnarray}
where in the last step we have used the formula
\begin{equation} \label{sintegral}
    \int_{0}^{\infty} \frac{ds}{s^{3/2}} e^{-s A -B/s} = \frac{e^{-2 \sqrt{AB}} \sqrt{\pi}}{\sqrt{B}}.
\end{equation}
Defining $t=4\pi i R A_4 + 2\pi R \sigma$ as the four-dimensional scalar we therefore see that the correction to the four-dimensional gauge-coupling is
\begin{equation}
    \frac{1}{g_4(t)^2} = K \sum_{n \neq 0} \frac{1}{|n|}e^{-4\pi i n R A_4 - 2\pi |n| R \sigma}
                       =  2 \textrm{Re}\left[\log(1-e^{-t})\right],
\end{equation}
with $K$ an overall normalization constant which is not important for our discussion.
As noted in \cite{Lawrence:1997jr} this is the effective perturbative gauge-coupling of $N=2,~D=4$ gauge theory with a cutoff $\Lambda_{UV} = \frac{1}{R}$.
If we take the limit $R \rightarrow 0$, the UV-cutoff disappears, and $t \rightarrow 0$ as well. As noted in \cite{Lawrence:1997jr,Katz:1996fh} we obtain the well-known result for the one-loop corrected purely four-dimensional gauge theory:
\begin{equation}
    \frac{1}{g_4(t)^2} \stackrel{R\rightarrow 0}{\sim} 2 \textrm{Re}\left[\log(t)\right].
\end{equation}

\subsubsection*{From 5D to 3D}

In going from $5D$ to $3D$ the five-dimensional vector-field splits as follows
\begin{equation}
    A_{\hmu} \rightarrow \left\{A_{\mu}, A_3, A_4\right\}~.
\end{equation}
$A_3$ and $A_4$ become scalars $\varphi_1$ and $\varphi_2$ in $3D$ defined in (\ref{phidef}). Dualizing the three-dimensional
vector-field $A_{\mu}$ to the scalar $\lambda$ gives, together with the $\varphi_i$ and $\sigma$, the bosonic sector of a
hypermultiplet.

In the remainder of this section, we compute the quantum corrected gauge kinetic coupling which corresponds to the quantum version of the function
$a(\sigma)$ in (\ref{Lclass}), i.e.
\begin{equation}
    a_{classical}(\sigma) \longrightarrow a_{quantum}(\sigma,\varphi_1,\varphi_2).
\end{equation}
This function cannot depend on $\lambda$, since in perturbation theory, the shift symmetry in $\lambda$ is preserved. Instanton effects break this
isometry, which we discuss in section \ref{instantons}.
In perturbation theory this function gets contributions from integrating out the Kaluza-Klein tower of
the massive vector-multiplet given by $(A^{\pm}_{\hmu}, \lambda^{\pm})$. We will compute it at one-loop by integrating out the massive fields. Going through steps similar to what we did from 5D to 4D we arrive at the following propagator for the massive fermions,
\begin{equation}
    \frac{1}{\Gamma^{\mu} k_{\mu} + \Gamma^3 (n_1 + \frac{\varphi_1}{2\pi}) + \Gamma^4 (n_2 + \frac{\varphi_2}{2\pi}) + \sigma}~,
\end{equation}
where the $n_i$ denote the Kaluza-Klein modes of the $T^2$ and $\sigma$ is the vev of the Coulomb branch parameter.
By using the anti-commutation relation
\begin{equation}
    \{\Gamma^{\hmu},\Gamma^{\hnu}\} = 2 g^{\hmu \hnu},
\end{equation}
where $g^{\hmu \hnu}$ is given in (\ref{5dmetric}), we can compute the one-loop amplitude with two propagators and two
gauge boson insertions. It leads to the gauge kinetic function at one-loop, and after dualizing the photon into a scalar, one obtains the hyperk\"ahler
metric (\ref{hkmetric}), with function $U$ given by\footnote{We exclude the zero-mode in the sum as it corresponds to the decompactification limit. To this respect we introduce a primed domain ${\IZ^2}^{\prime}$ which corresponds to $\IZ^2-\{0\}$.}
\begin{equation} \label{uoneloop}
    U_{1-loop} = K \sum_{\vec{n}\in{\IZ^2}^\prime} \int \frac{d^3 k}{(2\pi)^3} \frac{1}{\left[k^2 + \sigma^2
                + g^{ij}(n_i+\frac{\varphi_i}{2\pi}) (n_j + \frac{\varphi_j}{2\pi})\right]^2}~,
\end{equation}
with $K$ a normalization constant. The resultin function $U_{1-loop}$ in (\ref{uoneloop}) is obtained by integrating out massive fermions. However,
one also needs to integrate out the massive gauge bosons, to preserve supersymmetry. This can be done along the same lines as in \cite{Dorey:1997ij}.
The formula for $U_{1-loop}$ in (\ref{uoneloop}) then remains the same, only the coefficient $K$ changes. We will fix $K$ later on
by taking a specific limit. Using the identity
\begin{equation}
    \frac{1}{\left[k^2+\sigma^2 + g^{ij}(n_i+\frac{\varphi_i}{2\pi}) (n_j + \frac{\varphi_j}{2\pi})\right]^2} =
    \int_{0}^{\infty}ds s e^{-s \left[k^2+\sigma^2 + g^{ij}(n_i+\frac{\varphi_i}{2\pi}) (n_j + \frac{\varphi_j}{2\pi})\right]},
\end{equation}
and integrating over $k$ we obtain
\begin{equation}
    U_{1-loop} = \frac{K}{8 \pi^{3/2}} \sum_{\vec{n} \in {\IZ^2}^{\prime}}
        \int_{0}^{\infty} \frac{ds}{s^{1/2}} e^{-s\left[\sigma^2 + g^{ij}(n_i+\frac{\varphi_i}{2\pi}) (n_j + \frac{\varphi_j}{2\pi})\right]}~.
\end{equation}
As a next step we Poisson resum in $\vec{n}$:
\begin{equation}
    \sum_{\vec{n} \in {\IZ^2}^{\prime}} e^{-s g^{ij}(n_i + \frac{\varphi_i}{2\pi})(n_j+\frac{\varphi_j}{2\pi})}
    = \sum_{\vec{m} \in {\IZ^2}^{\prime}} e^{-2 \pi i m^i \frac{\varphi_i}{2\pi}} e^{-\frac{\pi^2}{s} g_{ij}m^i m^j} \sqrt{\textrm{det} g_{ij}} \frac{\pi}{s},
\end{equation}
yielding
\begin{eqnarray}
    U_{1-loop}
      & = & \frac{K}{8 \sqrt{\pi}} \sqrt{\textrm{det}g_{ij}}\int_{0}^{\infty} \frac{ds}{s^{3/2}} \sum_{\vec{m} \in {\IZ^2}^{\prime}}
            e^{-i m^i \varphi_i} e^{-\pi^2 g_{ij} m^i m^j/s - s \sigma^2} \nonumber \\
    ~ & = & \frac{K}{8 \pi^{1/2}} \sqrt{\textrm{det}g_{ij}} \sum_{\vec{m} \in {\IZ^2}^{\prime}} \frac{1}{\sqrt{g_{ij} m^i m^j}} e^{-i m^i \varphi_i
            - 2\pi \sigma \sqrt{g_{ij} m^i m^j}} \nonumber \\
    ~ & = & \frac{K}{8 \pi^{1/2}} \cV \sum_{n,m \in \IZ} \frac{1}{|m \tau + n|(\frac{\cV}{\tau_2})^{1/2}}
            e^{-i (m \varphi_1 + n \varphi_2) - 2 \pi \sigma |m \tau + n| (\frac{\cV}{\tau_2})^{1/2}} \label{Ufunction},
\end{eqnarray}
where in the second step we have again used formula (\ref{sintegral}). Indeed we see that the continuous shifts in $\varphi_1$ and $\varphi_2$ which were
isometries of the classical theory get broken to the discrete subgroup of shifts by integer multiples of $2\pi$.

Taking various limits of the above function $U$ in the parameter-space specified by $\tau$ we can establish contact with various metrics in the literature.
We will discuss two specific cases here. To this respect we go to the limit of a rectangular torus $T^2$, i.e. take
\begin{equation}
    \tau_1 = 0, \quad \tau_2 = \frac{R_1}{R_2}, \quad \cV = R_1 R_2.
\end{equation}
The first limit we take is $R_1 \rightarrow 0$ and $R_2 \rightarrow 0$. This is the limit in which the gauge theory in three-dimensional. It was studied in \cite{Seiberg:1996nz,Dorey:1997ij} where the metric on the Coulomb branch is the Atiyah-Hitchin metric. At the perturbative level, it becomes the Taub-Nut metric with
\begin{equation}\label{TN}
  U(\vec{x}) = \frac{1}{2g_{3,0}^2} - \frac{1}{4\pi|\vec{x}|}~.
\end{equation}
By comparing the second term with (\ref{uoneloop}) in the limit $R_1,~R_2 \rightarrow 0$, one then fixes the normalization constant $K=2$.
Next, take the limit $R_2 \rightarrow 0$ and $R_1$ arbitrary where we note that $\varphi_2 \rightarrow 0$ as well. This is the limit of
four-dimensional gauge theory on $\IR^3 \times S_{R_1}^1$.
In order to do this we split the double sum into the contributions $m = 0$ and $m \neq 0,~n \in \IZ$.:
\begin{equation}
    U_{1-loop} \sim \sum_{n \neq 0} \frac{1}{|n|} e^{-2\pi |n| R_2 \sigma - i n \varphi_2}
        + \sum_{m \neq 0} \sum_{n \in \IZ} \frac{1}{|m\tau + n|} e^{-2\pi |m\tau + n| R_2 \sigma - i m \varphi_1 - i n \varphi_2}.
\end{equation}
The first term is familiar from the 5D to 4D story and can be summed up to give
\begin{equation} \label{sslimit1}
    \sum_{n \neq 0} \frac{1}{|n|} e^{-2\pi |n| R_2 \sigma - i n \varphi_2} \stackrel{R_2 \rightarrow 0}{\simeq} \log(t \bar{t}),
\end{equation}
where we have identified $2\pi (R_2 \sigma + \frac{\varphi_2}{2\pi})$ with $t$ and ignored subleading corrections.
The second term can be Poisson resummed in $n$ and one arrives at
\begin{eqnarray}
    ~ & ~ & \sum_{m \neq 0} \sum_{n \in \IZ} \frac{1}{|m\tau + n|} e^{-2\pi |m \tau + n|R_2 \sigma - i m \varphi_1 - i n \varphi_2} \nonumber \\
    ~ & = & 2 \sum_{m\neq 0}\sum_{n \in \IZ} K_0\left(2\pi \frac{R_1}{R_2}|m|\sqrt{(R_2 \sigma)^2 + (\frac{\varphi_2}{2\pi} + n)^2}\right)
            e^{i m \varphi_1} \nonumber \\
    ~ & \simeq & 2 \sum_{m \neq 0} K_0\left(\frac{R_1}{R_2}|m t|\right)e^{i m \varphi_1}, \label{sslimit2}
\end{eqnarray}
where in the last line we have used that in the limit we are taking the sum localizes on $n=0$. The sum of (\ref{sslimit1}) and
(\ref{sslimit2}) agrees precisely with the result of Seiberg and Shenker \cite{Seiberg:1996ns} or Ooguri and Vafa \cite{Ooguri:1996me}
(see also \cite{Gaiotto:2008cd,deWit:1997vg}) if we identify $R_2$ with the string coupling. This brings us naturally to the M-theory interpretation of our results where $R_2$ is identified with the M-theory circle whose radius is $l_s g_s$.
In the next section we will elaborate on this picture.

\section{Embedding into string theory}

This section starts by embedding the five-dimensional gauge theory in a M-theory compactification. We then proceed to make contact with type IIA by
compactifying on a circle and to type IIB by compactifying on a torus. At each step we provide a dictionary between field theory quantities
and their string-/M-theory representations.

\subsection{5D supersymmetric gauge theories from M-theory}

Five-dimensional $SU(2)$ gauge theories in 5D with $N=1$ supersymmetry can be obtained by compactifying M-theory on a Calabi-Yau
manifold $X$ which is locally the total space of the canonical line bundle $\cO(K_P) \rightarrow P$ where $P$ is a del Pezzo surface.
Let us review the basic facts of this construction where we will refer to the references \cite{Morrison:1996xf,Douglas:1996xp} and to the
appendix of \cite{Alim:2010cf} for further details.

By definition, del Pezzo surfaces have positive anti-canonical class $D = -K_P$, i.e. $D^2 >0$, which makes them rigid inside the Calabi-Yau
and satisfy $D \cdot C > 0$ for any irreducible curve $C$ inside $P$.
Such surfaces are either $\IB_n$, which are blow-ups of $\IP^2$ in $n \leq 8$ points or blow-ups of $\IF_0 = \IP^1 \times \IP^1$. In the case of
$\IB_n$ the anti-canonical class is given by
\begin{equation} \label{dPclass}
    -K_{\IB_n} = 3h - \sum_{i=1}^n e_i,
\end{equation}
where $h$ is the homology class generated by the
hyperplane class $h$ of $\IP^2$ and the $e_i$ are the exceptional divisors of the blow-ups. The non-vanishing intersections
are
\begin{equation} \label{dPint}
    h^2 = 1 = - e_i^2.
\end{equation}
More intuitively, $\IB_n$ can be viewed as a fiber space over $\IP^1$ where the generic fiber is $\IP^1$,
but over $n-1$ points the fiber has two $\IP^1$'s intersecting at a point, i.e. the geometry of a resolved $A_2$ singularity.
Let us explain now how this picture reproduces the field theory description of section \ref{U1gaugeth}. In order to do that we
consider the limit in which the del Pezzo surface has a large base parameterized by $\phi_B$ and a small fiber parameterized by $\phi_f$.
Then $\phi_f$ corresponds to the Coulomb branch parameter of the field theory and $\phi_B$ will be identified with the bare coupling constant
of $SU(2)$, i.e. we have
\begin{equation} \label{cygaugethmap}
    \frac{1}{g_{5,0}^2} = \phi_B, \quad \phi = \phi_f.
\end{equation}
Furthermore, for a del Pezzo surface which is $\IB_n$, there are $n-1$ hypermultiplets which are doublets of $SU(2)$, that is $N_f = n-1$.
Going to the Coulomb branch of $SU(2)$ corresponds to resolving the generic $A_1$ singularity of the fiber (corresponding to $\phi_f=0$) and resolving the $n-1$ $A_2$ singular fibers.
The full field dependent gauge coupling is computed in terms of the intersection theory of $X$:
\begin{equation}
    \frac{1}{g(\phi)^2} = \frac{\p^2 \cF}{\p \phi^2} = 2 [\IB_n] \cdot [\IB_n] \cdot (\phi_B H_0 + \sum_{i=1}^{n-1} m_i H_i + \phi [\IB_n]),
\end{equation}
where $[\IB_n]$ is the Poincare dual two-form to the surface $\IB_n$, the dot-product corresponds to the wedge-product of two-forms, and the dot-product of three two-forms also implies integration over the Calabi-Yau. Furthermore, $H_0$, $H_i$ and $[\IB_n]$ form a basis for the subspace of $H^{1,1}(X)$ which is not orthogonal to $[\IB_n] \cdot [\IB_n]$ and the factor $2$ is
a normalization factor. In order to see how equation (\ref{gaugecoupling}) is reproduced we go to a sub-wedge of the K\"ahler cone where
$\phi > 0, ~m_i>0$ and $\phi > m_i$. For this parameter-space (\ref{gaugecoupling}) simplifies to
\begin{equation}
    \frac{1}{g(\phi)^2} = \frac{1}{g_{5,0}^2} + (16 - 2 N_f) \phi.
\end{equation}
Assuming that $[\IB_n] \cdot [\IB_n] \cdot H_0 = \frac{1}{2}$ and $[\IB_n] \cdot [\IB_n] \cdot H_i = 0$\footnote{This can be always realized for del Pezzo surfaces with the above parameter values.} we are left with the computation of $[\IB_n] \cdot [\IB_n] \cdot [\IB_n]$. But this is the self-intersection number
of the del Pezzo surface and can be computed using (\ref{dPclass}) and (\ref{dPint}) to give
\begin{equation}
    [-K_{\IB_n}] \cdot [-K_{\IB_n}] = 9 - n = 8 - N_f,
\end{equation}
which reproduces exactly the quantum field theory computation. Actually, equation (\ref{gaugecoupling}) is reproduced for all values of the parameters $\phi$
and $m_i$, i.e. it is valid on the union of all K\"ahler cones, as was shown in \cite{Morrison:1996xf}.

\subsubsection*{The geometry behind the gauge theory}
\label{ourCY}

In this paper we are dealing with $SU(2)$ gauge theory with no hypermultiplets in the douplet of $SU(2)$, i.e. we have the situation $N_f = 0$.
This corresponds to a Calabi-Yau containing $\IF_0$ or $\IB_1 = \IF_1$, with the two choices leading to different topological sectors of the theory \cite{Douglas:1996xp}, where our particular gauge theory corresponds to the choice $\IF_0$. One can explicitly construct a Calabi-Yau which contains this surface and then take the local limit to obtain the gauge theory. For that, let $X$ be an elliptic fibration over $\mathbb{F}_0$ given by a generic section of the anti-canonical bundle of the toric ambient space specified by the following vertices
\begin{equation*}
\begin{split}
D_0=(0,0,0,0), &\quad D_1=(0,0,0,1), \quad D_2=(0,0,1,0),\\
D_3=(0,0,-2,-3),& \quad D_4=(0,-1,-2,-3), \quad D_5=(0,1,-2,-3),\\
D_6=(1,0,-2,-3),& \quad D_7=(-1,0,-2,-3).
\end{split}
\end{equation*}
One finds large volume phases with the following Mori-vectors
\begin{center}
\begin{tabular}{c|c|ccccccc|c}
 & $D_0$ & $D_1$ & $D_2$ & $D_3$& $D_4$& $D_5$& $D_6$& $D_7$& \\
\hline
$l^1=$ & $-6$ & 3 & 2 & 1 & 0 & 0 &0 &0 & $C^E$ \\
$l^2=$ & 0 & 0 & 0 & $-2$ & 1 & 1 &0 &0 & $C^f$ \\
$l^3=$ & 0 & 0 & 0 & $-2$ & $0$ & 0 &1 &1 & $C^B$. \\
\end{tabular}
\end{center}
Here, the $C^A$ with $A=E,f,B$ correspond to a basis of  $H_2(X,\IZ)$, where $C^E$ corresponds to the elliptic fiber, $C^f$ the $\IP^1$-fiber of $\IF_0$ and
$C^B$ represents the $\IP^1$-base of $\IF_0$. Let $K_A$ be a Poincar\'e dual basis of the Chow group of linearly independent divisors of $X$, i.e.~$\int_{C^A}K_B=\delta^A_B$. The divisors $D_i=l_i^AK_A$ have intersections with the cycles $C^A$ given by $D_i.C^A=l_i^A$. We then have the following non-vanishing intersections of the divisors,
\begin{equation}
    K_E\cdot K_f \cdot K_B=1,\quad K_E^2 \cdot K_f=2,\quad K_E^2 \cdot K_B=2,\quad K_E^3=8.
\end{equation}
The divisor giving the Hirzebruch surface inside the Calabi-Yau manifold corresponds to
\begin{equation}
    [\mathbb{F}_0]=D_3=K_E-2K_f-2K_B.
\end{equation}
We parameterize the K\"ahler-form of this Calabi-Yau as follows
\begin{equation}
    J = \phi_E K_E - \frac{\phi_B}{4} K_B + \phi_f [\IF_0].
\end{equation}
Thus, the gauge coupling of the gauge theory becomes
\begin{equation}
    \frac{\p^2 \cF}{\p \phi_f^2} = 2 [\IF_0] \cdot [\IF_0] \cdot (-\frac{\phi_B}{4} K_B + \phi_f [\IF_0]) = \phi_B + 16 \phi_f.
\end{equation}
The K\"ahler cone of $X$ is determined by the conditions
\begin{eqnarray}
    0 & < & (\phi_E K_E - \frac{\phi_B}{4} K_B + \phi_f [\IF_0])\cdot (-C^f) = 2 \phi_f \\
    0 & < & (\phi_E K_E - \frac{\phi_B}{4} K_B + \phi_f [\IF_0])\cdot (-C^B) = \frac{\phi_B}{4} + 2 \phi_f \sim \frac{\phi_B}{4} \\
    0 & < & (\phi_E K_E - \frac{\phi_B}{4} K_B + \phi_f [\IF_0])\cdot C^E = \phi_E + \phi_f,
\end{eqnarray}
where in the second line we have used that we are in the weak coupling regime of the gauge theory and thus $\phi_B \gg \phi_f$.
However, we will not necessarily restrict ourselves to this limit. The local model is now obtained by scaling $\phi_E$ to infinity.

\subsubsection*{BPS spectrum}
\label{Mtheoryspectrum}

BPS states are obtained by wrapping M2-branes and M5-branes on various holomorphic subspaces of our local Calabi-Yau. Wrapping
M2 branes $n$ times around the fiber $\IP_f^1$ and $m$ times around the base $\IP_B^1$ we arrive at hypermultiplets in five dimensions
whose degeneracies are given by the Gopakumar-Vafa invariants $n^{(0)}_{n,m}$. Let us now come to the M5-branes.
These can wrap the divisor $\IF_0$ and give rise to BPS string states in five dimensions. The tension of these strings is given
by the volume of the divisor $\IF_0$ inside the Calabi-Yau:
\begin{eqnarray}
    \textrm{Vol}(\IF_0) & = & \int_{\IF_0} J \wedge J \nonumber \\
    ~                   & = & [\IF_0] \cdot (-\frac{\phi_B}{4} K_B  + \phi_f [\IF_0]) \cdot (-\frac{\phi_B}{4} K_B + \phi_f [\IF_0]) \nonumber \\
    ~                   & = & \phi_B \phi_f + 8 \phi_f^2.
\end{eqnarray}
Notice that under the identifications (\ref{cygaugethmap}) \textit{this volume becomes identical to the tension of the
monopole string} of the five-dimensional gauge theory given in (\ref{stringtension})! Therefore we can identify the two strings and
the dictionary between the five-dimensional field theory discussed in section \ref{5Dfieldth} and the M-theory compactification is as follows:

%\vspace{5mm}
\begin{table}[h]
\begin{center}
\begin{tabular}{|c|c|c|}
    \hline
    ~                       & field theory       & M-theory     \\
    \hline
    coupling                & $\frac{1}{g_{5,0}^2}$  & $\phi_B$     \\
    moduli                  & $\phi$             & $\phi_f$     \\
    \hline
    \multirow{3}{*}{states} & W-bosons           & $M2/\IP^1_f$ \\
                            & dyonic instantons  & $M2/\IP^1_B$ \\
                            & strings            & $M5/\IF_0$   \\
    \hline
\end{tabular}
\end{center}
\caption{Dictionary between five-dimensional field theory and M-theory.}
\end{table}
%\vspace{5mm}

These states are charged under the M-theory three-form $C_3$, its dual and their reduction to $5$ dimensions\footnote{Note that there is also a gauge field $A^B_{\hmu}$ from the reduction of the trhee-form along $\IP^1_B$. However, this vectorfield together with its whole multiplet is frozen and thus disappears in the gauge theory limit.}
\begin{equation}
    2 A^f_{\hmu} = \int_{\IP^1_f} C_3.
\end{equation}
In three dimensions the reduction of $C_3$ gives rise to $\varphi_1$ and $\varphi_2$ as follows:
\begin{equation}
    \varphi_1 = \int_{\IP^1_f \times S^1_A} C_3, \quad \varphi_2 = \int_{\IP^1_f \times S^1_B} C_3\,
\end{equation}
where $S^1_A$ and $S^1_B$ form a homology-basis of $T^2$. Thus M2-instantons are charged under the scalars $\varphi_i$. The M5-branes are charged under
the dual of $C_3$:
\begin{equation}
    \ast d C_3 = d C_6, \quad B_{\hmu \hnu} = \int_{\IF_0} C_6.
\end{equation}
In three dimensions the dictionary is
\begin{equation}
    \lambda = \int_{\IF_0 \times T^2} C_6.
\end{equation}
Thus M5-instantons are charged under $\lambda$.

\subsection{Type IIA viewpoint}
\label{TypeIIA}

In the previous subsection we have realized five-dimensional gauge theories by compactifying M theory on a Calabi-Yau. However, going to four dimensions,
gauge theories can be obtained by compactifying type IIA string theory on a Calabi-Yau manifold and taking the local limit as was
shown in \cite{Katz:1996fh}. How do the two pictures fit together? In \cite{Witten:1996qb}, this question was answered by noting that a four-dimensional field theory can be obtained by compactifying the five-dimensional one on a circle  $S^1$. When sending the radius $R$ of the $S^1$ to infinity, one gets M-theory
on $\IR^5 \times X$. On the other hand sending $R$ to zero gives Type IIA on $\IR^4 \times X$. Furthermore, the relation between the M-theory
metric $g_M$ on $\IR^4\ \times X$ and the Type IIA metric $g_{IIA}$ on $\IR^4 \times X$ is
\begin{equation}
    g_{M} = \frac{g_{IIA}}{T^{1/3} R}\ ,
\end{equation}
where $T$ is the two-brane tension. We have also encountered this relation in our computation from section \ref{5d4d} where we found that the
relation between the complex modulus $t_f$ of the four-dimensional theory and the real five-dimensional field $\phi_f$ is
\begin{equation}
    t_f = 2 \pi i R(2 A_{4,f} + i \phi_f)\ ,
\end{equation}
with $A_{4,f}$ being the fifth component of the five-dimensional vector-field. This means that when sending $R \rightarrow \infty$ while keeping $g_M$ or equivalently $\phi_f$ fixed, then $t_f$ is going to infinity. Therefore, M-theory in five dimensions only "sees" the region at infinity in the Type IIA
moduli space.

In this picture, the BPS states are obtained by wrapping D2-branes on holomorphic curves inside the Calabi-Yau. Moreover, for $SU(2)$ gauge theories,
W-bosons are obtained by wrapping D2-branes on $\IP^1_f$ and their mass is given by
\begin{equation}
    M_{W^{\pm}} = \frac{|t_f|}{R},
\end{equation}
where we have identified the radius of the M-theory circle with the string coupling, more precisely we have $R = l_s g_s$. This way taking the $R \rightarrow 0$
limit the W-bosons will have finite mass and the limit corresponds to the Seiberg-Witten limit taken in \cite{Lawrence:1997jr,Katz:1996fh}.
Also as was shown in \cite{Dimofte:2009tm} magnetically charged states are obtained by wrapping a D4-brane on the Hirzebruch surface $\IF_0$. Thus we arrive at the map between four-dimensional gauge theory and type IIA string theory presented in table 2.

\begin{table}[h]
\begin{center}
\begin{tabular}{|c|c|c|}
    \hline
    ~                       & field theory          & Type IIA \\
    \hline
    coupling                & $\frac{4\pi}{g_{4,0}^2} + 2\pi i \theta$     & $t_B = 2\pi R \phi_B + i \int_{\IP^1_B} B^{NS}_2$ \\
    moduli                  & $t$                                           & $t_f = 2\pi R \phi_f + i \int_{\IP^1_f} B^{NS}_2$ \\
    \hline
    \multirow{3}{*}{states/instantons} & W-bosons                                      & $D2/\IP^1_f$ \\
                            & Instantons                                    & Euclidean $F$-string/$\IP^1_B$ \\
                            & Magnetic monopoles                             & $D4/\IF_0$ \\
    \hline
\end{tabular}
\end{center}
\caption{Dictionary between four-dimensional field theory and Type IIA.}
\end{table}

\subsection{Type IIB viewpoint and embedding into Quaternion-K\"ahler geometry}
\label{IIBviewpoint}

Five-dimensional gauge theory on $\IR^3\times T^2$ can be engineered from compactifications of M-theory on $CY \times T^2$. The latter has a type IIB dual. To make this more explicit, we take one of the
cycles of the $T^2$ to correspond to the M-theory circle and the other to a space-time circle. For the ease of explanation, we restrict
ourselves to a rectangular torus with the M-theory circle having radius $R_2$ and the space-time circle $R_1$. Compactifying type
IIB on $CY \times S^1_{\frac{1}{R_1}}$ we see that the resulting theory is T-dual to our M-theory compactification. The situation is
depicted in figure 2.

\begin{figure}[h]
\label{figure:SeibergMap}
\center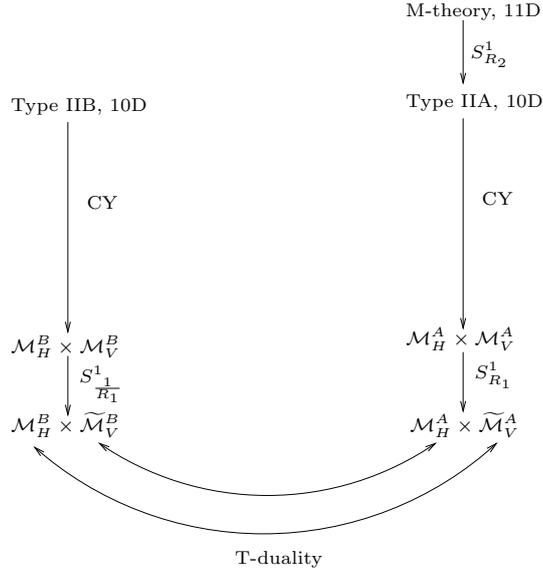
\caption{The map between type IIB and M-theory compactifications. ${\cal M}_H$ and ${\cal M}_V$ stand for the vector- and hypermultiplet moduli spaces that appear in the low-energy effective action. ${\widetilde {\cal M}}_V$ stands for the tensor multiplet moduli space, obtained from dimensional reduction of ${\cal M}_V$ to three dimensions.}
\end{figure}

As shown in the picture under T-duality the tensor multiplet moduli space of the type IIA-compactification to three-dimensions is exchanged
with the type IIB hypermultiplet moduli space in four dimensions. Therefore, we expect that the quantum corrections of the five-dimensional
gauge theory compactified on $T^2$ reproduce the instanton corrections of type IIB in the field theory limit.
In particular M2-branes (corresponding to W-bosons in the field theory) with Kaluza-Klein momentum along $S^1_{R_2}$ are mapped to world-sheet instantons
and M2-branes with Kaluza-Klein momentum along $S^1_{R_1}$ get mapped to D1-instantons in the type IIB picture after T-duality.

This suggests that we can obtain our hyperk\"ahler metric as a rigid limit from a 
quaternion-K\"ahler space obtained by compactifying type IIB string theory on a  Calabi-Yau manifold. The metric on this quaternionic hypermultiplet moduli space was worked out in \cite{RoblesLlana:2006is} starting from a generic Calabi-Yau. Due to the number of commuting isometries we have in our context, the hypermultiplets can be dualized to tensor multiplets, which makes it easier to compare to the tensor multiplets from the type IIA picture. Such tensor multiplet effective actions can be described by a single function ${\cal L}$ in the conformal calculus approach in which one adds an additional conformal compensating tensor multiplet (see e.g. \cite{RoblesLlana:2007ae} for more details within the present context). The second derivatives of this function, denoted by   $\cL_{xx}$, then determine the kinetic terms in the effective Lagrangian for the tensor multiplet scalars \cite{Hitchin:1986ea}, collectively denoted by $x$. In \cite{Saueressig:2007dr} it was  shown that the contribution from all world-sheet and D1-instanton corrections for a generic Calabi-Yau space, leads to the result
\begin{equation} \label{QKU}
     \cL_{xx} = \frac{1}{4\pi r^0} \sum_{k_a} n^{(0)}_{k_a} \sum_{m,n} \frac{1}{|m \tau + n|} e^{-2\pi k_a (|m\tau+n|t^a - i m c^a - i n b^a)}\ .
\end{equation}
In this expression, the dilaton-axion appears as
\begin{equation}
    \tau = \tau_1 + i \tau_2 = a + i e^{-\phi}.
\end{equation}
Furthermore, the $c^a$ are RR scalars and generate the theta-angle like terms for the D1-instantons, the $b^a$ correspond to the NS
$B$-field reduced along a basis of holomorphic two-cycles and the $t^a$ are reductions of the K\"ahler form along this basis. $r^0$ is the conformal compensator, which we further discuss below.

Upon dualizing the tensor multiplets back to hypermultiplets, one obtaines a hyperk\"ahler metric, and not a quaternion-K\"ahler metric. It is called the hyperk\"ahler cone in the terminology of \cite{de Wit:2001dj}, and it 
is due to the presence of the conformal compensator $r^0$. The resulting metric
is precisely of the form \eqref{hkmetric}, with $\cL_{xx}$ identified with $U$ \cite{Hitchin:1986ea}. After performing the superconformal quotient, in which the conformal compensator is eliminated, one obtaines a quaternion-K\"ahler metric of one (quaternionic) dimension less. This quotient can be done explicitly following the results of \cite{de Wit:2001dj}. However, for our analysis, we take the rigid limit {\it instead} of performing the superconformal quotient. In practice, this amounts to freezing the conformal compensator $r^0$ to a constant in  \eqref{QKU}. This breaks the scaling symmetry and  produces a hyperk\"ahler space of the form \eqref{hkmetric} of one (quaternionic) dimension less, since the conformal compensator multiplet is frozen. It has the same dimension as the quaternionic space, and contains the same coordinates, but it is hyperk\"ahler. In essence, we have taken a rigid limit\footnote{Geometrically, this freezing procedure
amounts to taking a hyperk\"ahler quotient on the hyperk\"ahler cone \cite{N&P}.}.

Next, we specify to the Calabi-Yau relevant for the gauge theory description given in section \ref{ourCY}. We have three geometric moduli:
\begin{equation}
    t^1 = \textrm{Vol}(C^E), \quad t^2 = \textrm{Vol}(\IP^1_f), \quad t^3 = \textrm{Vol}(\IP^1_B).
\end{equation}
Thus we see that the dimension of the quaternion-K\"ahler hypermultiplet moduli space is
\begin{equation}
    \textrm{dim}\,\cM_{H} = 3 \times 4 + 4 = 16,
\end{equation}
where the last $4$ comes from the multiplet to which the dilaton-axion belongs (the universal hypermultiplet).
Taking the limit
\begin{equation}\label{limit}
    t^1 \rightarrow \infty, \quad t^3 \rightarrow \infty, \quad \tau \textrm{~frozen},
\end{equation}
we restrict to a subspace of the hypermultiplets which has real dimension $4$. This restriction can best be done in the conformal approach described above. There, one has an off-shell description in terms of tensor multiplets. The limit \eqref{limit} can then be taken at the level of the superfields for the tensor multiplets, without breaking supersymmetry. One then ends up with two tensor multiplets, one of which is the compensator. Dualizing to hypermultiplets, one gets an eight-dimensional hyperk\"ahler manifold which still includes the compensator. Instead of performing the superconformal quotient to obtain the quaternion-K\"ahler manifold, we simply freeze the compensator to a constant. The end product is therefore a four-dimensional hyperk\"ahler space which is the rigid limit we are interested in.

Notice furthermore that in this limit the only Gopakumar-Vafa invariants which contribute are $n_{k_1=0,k_2,k_3=0}$.
As we have $n^{(0)}_{0,1,0} = -2$ and $n^{(0)}_{0,k,0} = 0~~\forall~~k > 1$ \cite{Katz:1996fh} we see that (\ref{QKU}) simplifies to
\begin{equation}
    \cL_{xx} = \frac{-2}{4\pi r^0} \sum_{m,n} \frac{1}{|m\tau + n|} e^{-2\pi(|m\tau+n|t^2 - i m c^2 - i n b^2)}.
\end{equation}
We now find that this function becomes identical to our function $U_{1-loop}$ up to overall normalization,
after the following field identifications
\begin{equation}
    t^2 = \sigma \left(\frac{\cV}{\tau_2}\right)^{1/2}, \quad 2 \pi c^2 = \varphi_1, \quad 2 \pi b^2 = \varphi_2\ .
\end{equation}
We can understand the identification $t^2 = \sigma (\frac{\cV}{\tau_2})^{1/2}$ if we go to the rectangular limit of the $T^2$.
In this case we have $(\frac{\cV}{\tau_2})^{1/2} = R_2$ and thus $t^2 = R_2 \sigma$ which is the correct identification
between M-theory and type II moduli as discussed in \ref{TypeIIA}. In this light the limit taken in (\ref{sslimit1}),(\ref{sslimit2})
corresponds to the conifold-limit taken in \cite{Saueressig:2007dr}.

The map between type IIB and the field theory on $\IR^3\times T^2$ can be extended even further. Note that the field theory contains
instantons by wrapping the string on $T^2$. Lifted to M-theory, these correspond to M5-branes wrapped around $\IF_0\times T^2$. Under
M-theory/IIA duality these M5-branes become D4-branes which under a further T-duality get mapped to D3-branes. Thus we see that
the string-instantons correspond to D3-instantons in type IIB.

In the following table we summarize the correspondence between type IIB string theory on our CY and M-theory on $CY\times T^2$ as well as
five-dimensional $N=1$ SYM on $\IR^3 \times T^2$.

\begin{table}[h]
\begin{center}
\begin{tabular}{|c|c|c|c|}
    \hline
    ~                                  & Type IIB on $CY$                  & M-theory on $CY \times T^2$              & field theory\\
    \hline
    coupling                           & $a + i e^{-\phi}$                 & $\tau$                                   & $\tau$ \\
    \hline
    \multirow{4}{*}{moduli}            & $\int_{\IP^1_f} C^{RR}_2$         & $\int_{\IP^1_f \times S^1_{R_1}} C_3$    & $\varphi_1$ \\
                                       & $\int_{\IP^1_f} B^{NS}_2$         & $\int_{\IP^1_f \times S^1_{R_2}} C_3$    & $\varphi_2$ \\
                                       & $\int_{\IF_0} C^{RR}_4$           & $\int_{\IF_0\times T^2} C_6$             & $\lambda$ \\
                                       & $t^2 (\frac{\cV}{\tau_2})^{-1/2}$ & $2\phi_f$                                & $\sigma$ \\
                                       \hline
    \multirow{3}{*}{states/instantons} & D1-instantons/$\IP^1_f$           & $M2/\IP^1_f\times S^1_{R_1}$             & $W$-boson/$S^1_{R_1}$\\
                                       & F1-instantons/$\IP^1_f$           & $M2/\IP^1_f\times S^1_{R_2}$             & $W$-boson/$S^1_{R_2}$\\
                                       & D3-instantons/$\IF_0$             & Euclidean $M5/\IF_0\times T^2$           & string/$T^2$\\

    \hline
\end{tabular}
\end{center}
\caption{Dictionary between type IIB and M-theory. The dyonic instanton can also be added; on the type IIB side, it corresponds to wrapping F1- and D1-strings over the base $\IP^1_B$.}
\end{table}

\section{Instanton corrections}
\label{instantons}

In this section, we discuss some aspects of the instanton corrections that correct the hyperk\"ahler metric describing the Coulomb branch of the five-dimensional gauge theory on $\IR^3\times T^2$. There are two kinds of instantons:  the dyonic instanton, whose worldline wraps a one-cycle in $T^2$, and the string instanton, whose worldsheet covers the entire $T^2$. We will focus on the latter, since it will elucidate some interesting new connection between the string instanton, the $M5$ brane theory compactified on the divisor $\IF_0$, and four-dimensional gauge theories on $\IF_0$.

We do not aim at computing these instanton corrections in full detail, as this is beyond the scope of this paper. The full hyperk\"ahler metric including these corrections would be an extension of the metric obtained in \cite{Gaiotto:2008cd}, which is best formulated in terms of twistor geometry. Instead, we focus here on the structure of the instanton induced correlation functions, in particular on the contribution from the zero-mode sector which involves the integration over the moduli space of the instanton. The contribution from the non-zero modes, involving the perturbation theory around the instanton, is not covered by our analysis. As in \cite{Dorey:1997ij,Chen:2010yr}, the correlation functions are four-fermi correlators, which yield corrections to the Riemann tensor of the hyperk\"ahler metric on the Coulomb branch.

\subsection{String Instantons}

The five-dimensional magnetic string, with tension $T$ and central charge $Z_m$ given by equations \eqref{tension-string} and \eqref{stringtension}, can wrap the torus $T^2$ to produce an instanton in three dimensions. The real part of the instanton action for this configuration is (we take $n_m$ and $\phi$ positive without loss of generality)
\begin{equation}
S_m={\cal V}Z_m=n_m\Big(\frac{\phi}{g_{3,0}^2}+\kappa {\cal V}\phi^2\Big)\ ,
\end{equation}
where ${\cal V}$ is the volume of the torus, and $g_{3,0}$ is the coupling constant of the three-dimensional field theory.
This coupling constant follows from the five-dimensional one by $g_{5,0}^2={\cal V}g_{3,0}^2$ and gets renormalized by a one-loop correction, which absorbs the second term inside the brackets. The imaginary part of the instanton action is given by $n_m \lambda$ where $\lambda$ is dual to the three-dimensional field strength induced by the monopole. Any instanton induced correlator, in our case a four-fermi correlator in the three-dimensional theory, is weighted by the exponent of minus the instanton action
\begin{equation} \label{instantonweight}
    e^{-\cV Z_m - i n_m \lambda}\ .
\end{equation}
Notice again that such instanton corrections break the shift isometry in $\lambda$. This implies that the hyperk\"ahler metric on the Coulomb branch is no longer of the type \eqref{hkmetric}, and one needs to resort to the twistor space techniques as used in \cite{Gaiotto:2008cd}.

To compute these instanton induced correlators, one needs to perform an integral over the finite dimensional moduli space of collective coordinates, ${\cal M}$. In a theory with additional compactified dimensions, the collective coordinates become functions of the coordinates of the extra dimensions, and the integral becomes a functional integral, as we explain in more detail below. This path integral is weighted by the Euclidean action describing the dynamics of the instanton, in our case the worldvolume theory on the string worldsheet wrapped around the torus. The path integral then becomes a partition function which appears as a factor in front of (\ref{instantonweight}). To make this more precise, let us clarify the situation by reminding that the BPS magnetic string is the uplift of the four-dimensional BPS magnetic monopole to five dimensions. This monopole, in turn, is the instanton in three-dimensions. Consider now first a four-dimensional gauge theory on $\IR^3\times S^1_\beta$, where the Euclidean time $\tau$ is compactified on a circle of radius $\beta$. At low energies (weak coupling), the dynamics on the moduli space ${\cal M}$ is described by a quantum mechanical path integral with an action for the collective coordinates of the form
\begin{equation}
S=\frac{1}{2}\int_0^\beta {\rm d}\tau \,\, g_{ij}(x)\dot{x}^i\dot{x}^j\ ,
\end{equation}
where $x^i$ denote the set of collective coordinates of the monopole, and $g_{ij}(x)$ is the metric on ${\cal M}$. For magnetic charge $n_m=2$, ${\cal M}$ factors into a part related to the center of mass motion, and a part describing the relative motion, which comes with the Atiyah-Hitchin hyperk\"ahler metric. For a single monopole, or for the center of mass part when $n_m>1$, we have that ${\cal M}=\IR^3\times S^1$,
and the collective coordinates are given by the three positions $\vec{X}$ in $\IR^3$, and an angle $\varphi \in [0,2\pi]$ that is induced by large gauge transformations. The Lagrangian is given by
\begin{equation}\label{mono-action}
L=\frac{M}{2}|\dot{\vec{X}}|^2+\frac{M}{2|\phi|^2}(\dot{\varphi})^2\ ,
\end{equation}
where $M$ is the mass of the monopole, and $|\phi|$ is the vev introduced in section 2. The motion along the compact circle with coordinate $\varphi$ is necessary to describe also the dyons, as the momentum along the $S^1$ is quantized and equal to the electric charge. See e.g. \cite{Chen:2010yr} and references therein for more explanation. From here on, we focus on the case $n_m=1$.

Now we consider again the five-dimensional theory. Since the string is the uplift of the monopole, its wordsheet action will also be the uplift. The mass of the monopole becomes the tension of the string, and includes the one-loop correction mentioned above. The worldsheet action for the string can be constructed by letting the collective coordinates of the monopole depend on both time and the coordinate of the extra dimension $y$ on which the monopole does not depend. Introducing coordinates $\sigma^\alpha=(\tau,y)$, the Euclidean worldsheet action of the magnetic string then takes the form of a Polyakov action on the torus. For a string with one unit of magnetic charge, the action is
\begin{equation}
S=\frac{T}{2} \int_{T^2}{\rm d}^2 \sigma \Big[ (\partial_\alpha \vec{X})\cdot (\partial^\alpha \vec{X})+\frac{1}{|\phi|^2} \partial_\alpha \varphi \partial^\alpha \varphi\Big]\ ,
\end{equation}
where $T$ is the tension of the string. This action reduces to \eqref{mono-action} when the moduli are independent of $y$.

The above actions need to be supplemented by fermions. Indeed, the BPS magnetic monopole preserves half of the supersymmetry, so four fermionic zero modes will be generated by acting with the broken supercharges. They can be denoted by $\xi^A_\alpha$, where $\alpha=1,2$ are spinor indices which descend from the two chiral spinors of the four-dimensional theory, labeled by $A=1,2$. Adding these to the action \eqref{mono-action}, one obtains a supersymmetric quantum mechanics on the moduli space. Uplifting these fermions to the two-dimensional worldsheet, one gets two spinors of the same chirality. All together, the worldsheet action of the instantonic string defines a $(0,4)$ conformal field theory with $3$ non-compact scalars and $1$ compact scalar. Redefining ${\tilde \varphi}=\varphi/\phi$ to get a canonically normalized scalar field, the periodicity then depends on the vev, i.e., $\tilde\varphi \in [0,2\pi/\phi]$. The Euclidean path integral for the zero-modes of the instanton string then becomes equal to the partition function for this $(0,4)$ conformal field theory. This statement we claim holds for any value of the magnetic charge. The contribution of BPS-states to the partition function is captured by the following generalized elliptic genus, which eventually appears as a factor in front of (\ref{instantonweight})\footnote{We are using here that the sigma-model has a CFT description. This is guaranteed by the fact that the metric on the moduli space ${\cal M}$ is hyperk\"ahler.}:
\begin{equation}\label{ellipticgenus}
    \cZ_{n_m}(\tau,T,\phi) = \textrm{Tr}\left[(-1)^F F^2 q^{L_0-\frac{c_L}{24}} \bar{q}^{\bar{L}_0-\frac{c_R}{24}}\right],
\end{equation}
where $q = e^{2\pi i \tau}$ and $\tau$ is the modulus of the torus. Furthermore, $F=2\bar{J}^3_0$ and $F^2$ has been inserted to soak up the fermion zero-modes and raises the anti-holomorphic modular weight by $2$ which makes the partition function a modular form of weight $(0,2)$ \cite{deBoer:2006vg}. Lastly, the eigenvalues of $L_0$ and ${\bar L}_0$ do depend on modulus $\phi$ and the tension of the string. 
For $n_m=1$, this partition function can easily be computed, and the result is
\begin{equation} \label{stringpartition}
{\cal Z}_{n_m=1}(\tau,T,\phi)= \left(\frac{T}{2\pi \tau_2}\right)^{\frac{3}{2}} \eta(\tau)^{-4} \sum_{n,m=-\infty}^{\infty} q^{\frac{1}{2\pi T} p_L^2}
                        \bar{q}^{\frac{1}{2\pi T}p_R^2},
\end{equation}
where $p_{L}$ and $p_{R}$ are given by
\begin{equation}
    p_L = \frac{n}{2}\phi + \frac{m}{2\phi} 2\pi T, \quad p_R = \frac{n}{2}\phi - \frac{m}{2\phi} 2\pi T.
\end{equation}
This result can be understood as follows. The first factor, involving $\tau_2^{-3/2}$ comes from integrating over the momenta of the three non-compact bosons. The second factor, being $\eta(\tau)^{-4}$, arises from summing over the oscillator modes of the three non-compact bosons and the compact one. Here, it is important to note that we do not have factors $\frac{1}{\bar{\eta}(\tau)}$ as the right-movers are kept on the ground state due to the BPS condition \cite{Gaiotto:2006wm,deBoer:2006vg}. Finally, the last factor arises from summing over the discrete momenta of the compact boson.

Sofar, we have only discussed field theory considerations. As mentioned in section 3, the magnetic string is realized in $M$-theory as the $M5$-brane wrapping the divisor $\IF_0$ inside the Calabi-Yau manifold. The worldvolume dynamics of the $M5$ brane is described by a six-dimensional $(0,2)$ CFT on $\IF_0 \times T^2$. Compactifying this theory on $\IF_0$, we obtain at low energies a two-dimensional $(0,4)$ CFT on $T^2$ whose partition function we claim to be the one for the magnetic string. For $n_m=1$, one can prove this by using recent work on $M5$-branes \cite{Alim:2010cf}. This is consistent with the fact that the tension of the string is equal to the volume of $\IF_0$, as explained in section \ref{Mtheoryspectrum}. 

On the other hand, one can give an alternative, dual, viewpoint by using a 2d/4d correspondence \cite{Minahan:1998vr}. 
To show this, we can also compactify the worldvolume theory of the $M5$-brane on $T^2$. Using the results of section 3, this theory is described in terms of $D3$-branes in type IIB. The worldvolume theory is the maximally supersymmetric  $N=4$ $U(r)$ SYM theory on $\IF_0$ where $r$ is the number M5-branes. One of the seminal approaches to such theories was the work of Vafa and Witten \cite{Vafa:1994tf} in which they performed a topological twist and turned the theories into topological field theories. Furthermore, as noted in \cite{Minahan:1998vr}, in a reduction of the M5-brane worldvolume theory, the complex structure $\tau$ of the torus $T^2$ becomes the complexified gauge coupling of the four-dimensional gauge theory:
\begin{equation}
    \tau = \frac{\theta}{2\pi} + \frac{4\pi i}{g_{YM}^2}\ .
\end{equation}
The $SL(2,\IZ)$ symmetry of the $T^2$ then descends to S-duality acting on the gauge coupling implying that the partition function of the topological theory is
a modular form if the gauge group is $U(r)$. The partition function for $U(r)$ topological SYM on a surface $P$ has the following expansion
\begin{equation}
    \cZ^{(r)}_{P}(\tau) = q^{-\frac{r\chi(P)}{24}} \sum_n \chi(\cM_n) q^n,
\end{equation}
where for simplicity we have neglected all anti-holomorphic dependence and refer to \cite{Vafa:1994tf} for details. In the above expansion $n$ is the instanton number, i.e.
\begin{equation}
    n = \frac{1}{8\pi^2} \int_{P} \textrm{Tr}~F \wedge F.
\end{equation}
Furthermore, $\cM_n$ is the instanton moduli space and $\chi(\cM_n)$ its Euler number. Using the elliptic genus representation of $\cZ^{(r)}_{P}$ \cite{Alim:2010cf,Manschot:2011dj} which descripes the partition function as a modular form of weight $(-\frac{3}{2},\frac{1}{2})$ admitting a theta-function decomposition, and restricting to $U(1)$ gauge theory on $\IF_0$, the result reads\footnote{We have set the Wilson line expectation value which appears in \cite{Alim:2010cf} to $z=\frac{1}{2}$.}
\begin{equation}
  \cZ^ {(1)}_{\IF_0}(\tau) = \frac{1}{\eta(\tau)^{\chi(\IF_0)}} \theta_{\Gamma^{1,1}_{\IF_0}}(\tau).
\end{equation}
The first factor is the generating function for rank 1 sheaves on $\IF_0$ and was computed by G\"ottsche \cite{Gottsche}. In our case
it is $\eta(\tau)^{-4}$ as $\chi(\IF_0)=4$. The second factor arises from summing over $U(1)$ fluxes through two-cycles of $\IF_0$.
It is the theta function of the cohomology lattice $H^2(\IF_0,\IZ)$ equipped with the intersection form as metric:
\begin{equation}
  \theta_{\Gamma^{1,1}_{\IF_0}}(\tau) = \sum_{k\in H^2(\IF_0,\IZ)} e^{2\pi i \tau \frac{k_+^2}{2}}e^{2\pi i \bar{\tau}\frac{k_-^2}{2}},
\end{equation}
where
\begin{equation}
  k_+^2 = \frac{(k\cdot J)^2}{J\cdot J}, \quad k_-^2 = k^2 - k_+^2,
\end{equation}
and $J$ is the K\"ahler form on $\IF_0$. As the lattice $H^2(\IF_0,\IZ)$ is even and self-dual one sees that this theta-function coincides with the sum over the discrete momenta of the compact boson in (\ref{stringpartition}), with the identifications $k_+=p_L/{\sqrt {\pi T}}$ and $k_-=p_R/{\sqrt {\pi T}}$.
Thus we have proven the following relation between the 2d and 4d partition functions:
\begin{equation}
  \cZ_{n_m=1} = \left(\frac{T}{2\pi\tau_2}\right)^{\frac{3}{2}}\cZ^{(1)}_{\IF_0}.
\end{equation}
The proportionality factor is generic, i.e. it is independent of the magnetic charge $n_m$, as for $n_m$ monopoles one can always factor out
$\tau_2^{-3/2}$ coming from the integration over the zero modes of the center of mass scalars.

\section{Outlook}

In this paper we have studied various aspects of the Coulomb branch of five-dimensional $N=1$ supersymmetric $SU(2)$ gauge theories compactified on $T^2$. Our analysis is not complete and much work needs to be done. We now list a few directions for future research.

First of all, the contribution of
the dyonic instanton  has been neglected in our analysis.
Taking these into account should lead to a consistent hyperk\"ahler metric which in the limit
$R_2 \rightarrow 0$ and large $R_1$ reduces to the structure described in \cite{Seiberg:1996nz}, i.e. a fibration of the Jacobian of
the Seiberg-Witten curve for pure $SU(2)$ over the $u$-plane. To be more precise, the complex structure of the elliptic fiber
gets corrected by four-dimensional instantons which in our case should be reproduced by contributions of the dyonic instanton.

A further non-perturbative state of the five-dimensional field theory is the magnetic string. As we have seen in the last section
its zero-mode partition function is closely related to the partition function of $N=4$ SYM on the Hirzebruch-surface $\IF_0$.
It would be very interesting to explore this correspondence for higher magnetic charge, i.e. $n_m > 1$. Also in this context, knowledge of
the non-zero mode contributions of the string instanton might shed some light on the connection to four-dimensional gauge theories and the analysis performed in \cite{Chen:2010yr}. 

Another interesting direction for extending our work is the addition of flavor to the
gauge theory. Geometrically this corresponds to blowing up the Hirzebruch surfaces $\IF_0$ and $\IF_1$ to del Pezzo surfaces. Here the
type IIB interpretation of section \ref{IIBviewpoint} should provide some hints (or even concrete conjectures) as what to expect for the perturbative and non-perturbative gauge theory computations. Furthermore, higher rank gauge groups yield higher dimensional hyperk\"ahler metrics on the Coulomb branch of the 5D theory, so this is an interesting extension as well.

 Last but not least
we think that a twistor description of the hyperk\"ahler  metric will provide a firmer ground for exploring the non-perturbative contributions to the
metric and to establish connection with the work done in \cite{Gaiotto:2008cd}.

\acknowledgments
We would like to thank Rajesh Gupta, Albrecht Klemm and Boris Pioline for interesting discussions. Furthermore, we acknowledge support from the Netherlands Organization for Scientific Research (NWO) under the VICI grant 
680-47-603.

%%%%%%%%%%%%%%%%%%%%%%%%%%%%%%%%%%%%%%%%%%%%%%%%%%%%%%%%%%%%%%%%%%%%%%%%%%%%%%%%%
%%%%%%%%%%%%%%%%%%%%%%%%%%%%%%%%%%%%%%%%%%%%%%%%%%%%%%%%%%%%%%%%%%%%%%%%%%%%%%%%%

\end{document}